\documentclass[preprint,aps,prd,groupedaddress,floatfix,%
nofootinbib]{revtex4}

\usepackage{dcolumn}
\usepackage{graphicx}
\usepackage{epsfig}
\usepackage{hyperref}
\usepackage{graphicx}
\usepackage{dcolumn}
\usepackage{amssymb,eucal}
\usepackage{longtable}
\usepackage{amsmath,amssymb,bm}
\usepackage{mathrsfs}

\newcommand{\mathd}{\mathrm{d}}
\newcommand{\mathi}{\mathrm{i}}
\newcommand{\mathe}{\mathrm{e}}

\newcommand\tr{{\rm tr }}

\def\tr{\mathop{\rm tr}\nolimits}

\newcommand{\bra}[1]{\langle #1 |}
\newcommand{\ket}[1]{| #1 \rangle}

\begin{document}
\title{Planar limit of 1D many-body system}
\author{Fen Zuo\footnote{Email: \textsf{zuofen@hust.edu.cn}}}
\affiliation{Huazhong University of Science and Technology, Wuhan 430074, China}
\author{Yi-Hong Gao\footnote{Email: \textsf{gaoyh@itp.ac.cn}}}
\affiliation{State Key Laboratory of Theoretical Physics, Institute of Theoretical Physics,
Chinese Academy of Sciences, P.O. Box 2735, Beijing 100190, China}

\begin{abstract}
We review one dimensional matrix theory and its variations, collective field theory and quantum phase space description. In the planar limit, these theories become classical and can be easily analyzed. With these descriptions, one dimensional interacting many-body system can be solved exactly when the particle number goes to infinity. As an example, bosonic and two-component fermionic systems with a $\delta$-function interaction are analyzed in detail.

\end{abstract}
 \maketitle

\section{Introduction}
The idea of large-$N$ limit was implicitly proposed in early 1950s, for different models in condensed matter physics. See~\cite{Chatterjee:1990se} for the early history and development. In 1974 't Hooft applied it to QCD~\cite{'tHooft:1973jz}, and reorganize the expansion of the theory in terms of $1/N$ and $\lambda\equiv g_{YM}^2 N$. The leading diagrams in $1/N$ can be drawn only on a plane/sphere, and termed planar diagrams. The large-$N$ limit with such a reorganized expansion is thus known as planar limit. The $1/N$-suppressed terms can only be drawn on surfaces with higher genus, which indicates some relation with the topological expansion of the string scattering amplitudes. For ${\mathcal N}=4$ Super Yang-Mills theory, the planar diagrams are described by the non-interacting IIB string theory on 5 dimensional anti-de Sitter spacetime multiplied by a 5 sphere~\cite{Maldacena:1997re}. They are given as an expansion at large 't Hooft coupling $\lambda$, with the leading contribution generated explicitly by the corresponding supergravity. Following 't Hooft, the planar expansion is extended to the scalar field theories, where the scalar is uplifted to a matrix field~\cite{Brezin:1977sv}. In one dimension, the planar limit of such a matrix theory is solved by a non-interacting fermionic system, with the fermion positions representing the matrix eigenvalues. To describe realistic many-body system at large-$N$, one could generalize the matrix theory to a theory of the density function of the particles, the collective field~\cite{Jevicki:1979mb}. Alternatively, one could use the phase space representation of the density operator, which is called $W_\infty$ coadjoint orbit or coherent state approach~\cite{Dhar:1992hr,Dhar:1993jc}. In~\cite{Yaffe:1981vf}, it is shown that in general the coadjoint orbit/coherent state approach captures the classical dynamics in the large-$N$ limit.

Despite the early development of large-$N$ limit in condensed matter physics, the planar expansion is not widely employed. In 2009, Ma and Yang
analyzed the $N$-dependence of the groundstate energy of 1D two-component fermionic system, with both a $\delta$-function interaction and a harmonic confining potential~\cite{Ma0926505}. They found that the energy scales as $N^2$ when the interaction is repulsive, and $-g^2N$ when the interaction strength $g\to -\infty$. They conjectured that the re-scaled energy $E/N^2$ should be a smooth function of $g/\sqrt{N}$. One may immediately recognize this as a planar limit, of a properly generalized matrix theory. In this paper we will show how this could be indeed realized.

We organize the paper in the following way. In the next section we review the 1D matrix theory, its planar limit, and the solution through non-interacting fermions. In section III the collective field theory is briefly shown, and applied to 1D interacting bosonic system. Then in section IV we use the phase space description to describe 1D interacting fermionic system. We take the $\delta$-function interaction as an example, and compare the results with those from other approaches when available. In the last section a short summary is given.

\section{Matrix theory, planar diagrams and 1D free fermions}
First we review the relation between the 1D hermite matrix theory and the free fermion system~\cite{Brezin:1977sv}.
 We will repeat the detailed procedure, in order to show the relation to the fermionic system clearly. For the $\varphi^4$ theory, the matrix Lagrangian is given as
\begin{equation}
{\mathcal L}=\tr (\partial_t M \partial_t M^\dagger)+\tr(MM^\dagger)+g_4\tr(MM^\dagger MM^\dagger)
\end{equation}
Following the proposal of 't Hooft~\cite{'tHooft:1973jz}, the propagators of the matrix field are represented with double lines. Consider a general connected vacuum diagram, which contains $P$ propagators , $V$ vertices and $I$ closed loops of internal index. Viewing the internal loop as a surface, one obtains the Euler relation
\begin{equation}
V-P+I=2-2H.
\end{equation}
Here $H$ is the number of holes of the Feynman diagram. Taking into account $2P=4V$, one can re-arrange the factors to obtain the weight coefficient for the diagram
\begin{equation}
g_4^V N^I=(g_4N)^VN^{2-2H}.
\end{equation}
Therefore if we take $g_4\propto 1/N$, only planar diagrams with $H=0$ remain in the large-$N$ limit. In other words, one only needs to keep those diagrams that can be drawn on a plane or sphere. Explicitly, one may write
\begin{equation}
\lim_{N\to \infty} \int \mathd ^{N^2}M \,\, \mathe ^{-\int {\mathcal L} \mathd t } =\mathe ^{-N^2 E (g_4)}, \label{eq.vacuum}
\end{equation}
where $E_0(g_4)$ is the sum of all connected vacuum diagram on a planar surface. Notice that the conclusion will not be changed even if the interaction is of power $p\ne 4$. In such a case the corresponding coupling constant scales as $g_p\propto N^{1-p/2}$. In particular, for the quadratic term the coefficient does not change with $N$. It simply reflects the difference between the propagator and the interaction.

In general, it is still not easy to sum all the planar diagrams. In the present 1D case, it turns out to be equivalent to finding the ground state energy of the corresponding Hamiltomian
\begin{eqnarray}
H&=&-\frac{1}{2}\Delta +V\nonumber\\
\Delta&=&\sum_i \frac{\partial^2}{\partial M_{ii}^2}+\frac{1}{2} \sum_ {i<j} \left(\frac{\partial^2}{\partial \mbox{Re} M_{ij}^2}+\frac{\partial^2}{\partial \mbox{Im} M_{ij}^2}\right)\nonumber\\
V&=&\frac{1}{2}\tr{M^2}+g_4\tr{M^4}.
\end{eqnarray}
That is, we only need to solve the equation
\begin{equation}
H \psi = N^2 E(g_4)\psi,
\end{equation}
and find the ground state wave function. The wave function should be symmetric under the U(N) rotation $M\to UMU^{-1}$, as required in the matrix theory. Alternatively, the ground state energy can be obtained from minimization of the action within the invariant wave function configuration
\begin{equation}
E(g_4)=\lim_{N\to \infty} \frac{1}{N^2} \min_\psi \frac{\int \mathd ^{N^2} M\, (\frac{1}{2}(\partial \psi)^2+V\psi^2)}{\int \mathd ^{N^2} M\, \psi^2}.
\end{equation}
Therefore it indeed corresponds to the sum of the connected vacuum diagram (\ref{eq.vacuum}). The integration over the angle part $U$ can be done trivially, leaving the integration over the eigenvalues $x_i$ of $M$. So the ground state energy becomes
\begin{equation}
E(g_4)= \lim_{N\to \infty} \frac{1}{N^2} \min_\psi \frac{\prod_i \mathd x_i \prod _{i<j} (x_i-x_j)^2\left[\frac{1}{2}\sum_i \left(\frac{\partial \psi}{\partial x_i}\right)^2+V(x_i)\psi^2\right]}{\prod_i \mathd x_i \prod _{i<j} (x_i-x_j)^2\psi^2}.\label{eq.E-diagonal}
\end{equation}
Defining
\begin{equation}
\tilde \psi(x_1,...,x_N)= \left\{\prod_{i<j} (x_i-x_j)\right\} \psi(x_1,...,x_N),\label{eq.B-F}
\end{equation}
one obtains the Schr\"{o}dinger equation
\begin{equation}
\sum_i \left(-\frac{1}{2}\frac{\partial^2}{\partial x_i^2} +\frac{1}{2} x_i^2 + g_4 x_i^4\right)~\tilde \psi=N^2 E \tilde \psi.
\end{equation}
From the definition (\ref{eq.B-F}) we see that the above equation describes the emotion of $N$ fermions in 1D. They are confined in a central potential, but do not interact with each other. In the large-$N$ limit, they behave semi-classically and simply fill all the states below the Fermi energy level $e_F$. Integrating out the momentum, the energy and the fermion number are then expressed as
\begin{eqnarray}
N^2 E(g_4)&=&N e_F -\int \frac{\mathd x}{3\pi} [2e_F-x^2-2g_4x^4]^{3/2}\,\theta (2e_F-x^2-2g_4x^4)\nonumber\\
N&=& \int \frac{\mathd x}{\pi} [2e_F-x^2-2g_4x^4]^{1/2}\,\theta(2e_F-x^2-2g_4x^4)
\end{eqnarray}
Clearly at lage $N$, $e_F\sim N$, and the range of $x$, as well as the density distribution of $x$, scales as $\sqrt{N}$. Re-scaling the parameters as $e_F=N \varepsilon$ and $x=\sqrt{N} u$, the above relations can be simplified
\begin{eqnarray}
E(\lambda)&=&\varepsilon -\int \frac{\mathd u}{3\pi} [2\varepsilon-u^2-2\lambda u^4]^{3/2}\,\theta (2\varepsilon-u^2-2\lambda u^4)\nonumber\\
1&=& \int \frac{\mathd u}{\pi} [2\varepsilon-u^2-2\lambda u^4]^{1/2}\,\theta(2\varepsilon-u^2-2\lambda  u^4),
\end{eqnarray}
where $\lambda\equiv g_4 N$.
The density in the new variable $u$ now reads
\begin{equation}
\rho(u)= \frac{1}{\pi} [2\varepsilon-u^2-2\lambda u^4]^{1/2}\,\theta(2\varepsilon-u^2-2\lambda u^4).\label{eq.rho-free}
\end{equation}

Taking $g_4=0$ one recovers the famous semi-circle law of Wigner~\cite{Mehta67}, after redefining $u$ to absorb $\varepsilon$.
When $g_4$ is non-vanishing the semi-circle is deformed. One may set the quadratic term to zero instead, then the distribution function is completely determined by the quartic term. Nevertheless, scaling of various quantities with $N$ is unchanged.

How to generalize the above discussion to bosons? How to introduce the interactions between the particles? It is not so obvious in the matrix formalism, although some progress could still be made. In the next two sections we will show how to deal with them in the so-called collective field theory and the quantum phase space description. Before plunging into details, we can first sketch the $N$-scaling when the interactions are present. Since the coefficient of quadratic term~(mass) does not change in the planar limit, the range of $x$ and the density function $\phi(x)$ in $x$ are always of order $\sqrt{N}$. Integration $\phi(x)$ over $x$ then gives the correct particle number $N$. As a result, for a system with the $\delta$-function interaction
\begin{equation}
\Delta H=g\sum_{i<j} \delta(x_i-x_j),
\end{equation}
the planar limit can be achieved when $g/\sqrt{N}$ is kept fixed. This is exactly the hypothesis made in \cite{Ma0926505} and further confirmed in \cite{Ma0926506,Ma1027501}. With the $N$-scaling of the coupling constant taking into account, we could also recover the large-$N$ behavior obtained for a power potential trap~\cite{Ma1027505}.

\section{Collective field theory and 1D interacting bosons}
First we ignore the problem of the statistics, and try to reexpress the $\delta$-function interaction in a matrix form. The answer is almost immediate
\begin{equation}
\Delta H=g\sum_{i<j} \delta(x_i-x_j)\sim \frac{g}{2}~\int \frac{\mathd k}{2\pi} ~\tr(\mathe ^{-\mathi k M} ) ~\tr (\mathe ^{\mathi k M}).
\end{equation}
However, the factor $\prod _{i<j} (x_i-x_j)^2$ from the matrix integration measure (\ref{eq.E-diagonal}) will force the contact two-body interaction to vanish. Therefore, it is convenient to leave the matrix formalism and deal directly with the collective field~\cite{Jevicki:1979mb}
\begin{equation}
\phi(x)=\int \frac{\mathd k}{2\pi} \mathe ^{\mathi kx}~\tr(\mathe^{-\mathi k M})=\sum_{i=1}^N \delta(x-x_i).
\end{equation}
It describes the density of the particles at a fixed position $x$. Now we want to use $\phi(x)$ to describe the system, instead of the positions $x_1,...,x_N$. Explicitly, one could restrict $x$ in some finite interval $-L/2\le x\le L/2$, and start with the Fourier modes $\phi_k$ of $\phi(x)$
\begin{equation}
\phi_k=\frac{1}{L}\int \mathd x \mathe ^{-\mathi k x} \phi(x)=\frac{1}{L} \tr(\mathe ^{-\mathi k M} ),
\end{equation}
with $k=2\pi n/L$ and $n$ takes integer numbers. Clearly $\phi(x)$ contains more than enough degrees of freedom than the positions, even for finite $L$. That means the Fouries modes $\phi_k$ may not be independent. However, at least in the large-$N$ limit this will not cause any problem.


Consider a bosonic system with a general Hamiltonian
\begin{equation}
H=\frac{1}{2}\sum_{i=1}^N p_i^2 +\frac{1}{2}\sum_{i\ne j}^N v(x_i,x_j)+\sum_{i=1}^N V(x_i). \label{eq.1D-Boson}
\end{equation}
We need to make the coordinate transformation from $x_i$ to $\phi_k$. In other words, we consider now the many-body wave function $\psi[x_i]$ as a composite function $\psi[\phi_k(x_i)]$, and rewrite $H$ in terms of $\phi_k$. In making the variable change one has to take care with the jacobian. Taking $L\to \infty$ in the end, the expression in terms of $\phi_k$ can be compactly expressed through the field $\phi(x)$. After a sightly long derivation one obtains the following Hamiltonian~\cite{Jevicki:1979mb,Das:1990kaa}
\begin{eqnarray}
H_\phi&=&\int \mathd x \left[ \frac{1}{2} \partial_x \pi \phi \partial_x \pi + \frac{\pi^2}{6}\phi^3(x)+\Delta V+V_0\right]\label{eq.H-phi}\\
\Delta V&=&\frac{1}{8}\int \frac{(\partial_x \phi)^2}{\phi}\mathd x-\frac{1}{2}\int \partial_x \phi(x) \left[\int \frac{\phi(y)}{(x-y)}\mathd y\right]\mathd x\nonumber\\
V_0&=& \int \bigg[ -\big(\mu_F-V(x)+v(x,x)\big)\phi(x)\bigg]\mathd x\nonumber\\
&&\quad\quad+\frac{1}{2}\int \mathd x \int \mathd y ~~\phi(x) v(x,y)\phi(y).
\end{eqnarray}
where $\pi(x)\equiv \frac{1}{\mathi}\frac{\delta}{\delta \phi(x)}$ is the conjugate field of $\phi(x)$, $\mu_F$ is a Lagrangian multiplier to ensure the constraint
\begin{equation}
\int \mathd x ~\phi(x)=N. \label{eq.Norm1}
\end{equation}
Notice that the first term in $\Delta V$ is the well-known Weizs\"{a}cker term~\cite{Weizsacker1935}, introduced to better describe the kinetic term in the Thomas-Fermi model. With the previous $N$-scaling arguments, one can easily find that $\Delta V$ is suppressed in the planar limit. This will be manifest if we rescale the density function and the coordinate as
\begin{equation}
x\equiv \sqrt{N} u,\quad \phi(x)\equiv \sqrt{N} \rho(u),\label{eq.N-scaling}
\end{equation}
and redefine the coefficients in the potential $V(x)$ and the interaction $v(x,y)$ accordingly. The re-scaled variables $u$ and $\rho$ will be independent of $N$. Only an overall factor $N^2$ remains in the Hamiltonian, as expected from the discussion in the previous section. Since the field behaves classically in the large-$N$ limit, we can determine the density function by minimizing the potential. Notice that although the kinetic term in the Hamiltonian is of leading order $N^2$, the contribution can be neglected when we consider the density to be stable and not evolving with time.

Let us apply the above formalism to an explicit example. We take the potential to be harmonic, and the interaction of the $\delta$-function form
\begin{equation}
V(x)=\frac{1}{2}\omega^2 x^2,\quad v(x,y)=g \delta(x-y).
\end{equation}
When the harmonic potential is absent, this gives the Lieb-Liniger model~\cite{Lieb:1963rt}, which could be solved analytically through the Bethe ansatz~\cite{Bethe1931}. In the planar limit $\omega$ remains unchanged, and $g$ behaves as $\sqrt{N}$. Therefore the relevant coupling will be $\alpha\equiv g/\sqrt{N}$. So the conjecture made in \cite{Ma0926505} is just the consequence of the planar limit. An equivalent parameter has been previously used in~\cite{Lieb:1963rt,Dunjko2001}. Actually one may identify the large-$N$ limit in~\cite{Lieb:1963rt} as the planar limit, with the density defined there scaling as $\sqrt{N}$. An immediate consequence of the above scaling (\ref{eq.N-scaling}) is, in each small interval $\mathd x$ with nonzero density, there will be as many as ${\mathcal O}(\sqrt{N})$ particles. One could apply a localized version of the Fredholm equation~\cite{Lieb:1963rt}, and find the local energy density and Gibbs energy density~\cite{Dunjko2001,Ma0926506,Ma1027501,Ma1027505}. These results can then be used as input to solve the remaining hydrodynamic equations in a potential.

With the new variable and field, the leading order potential reads
\begin{equation}
V_{eff} = N^2 \int \left\{ \frac{\pi^2}{6}\rho^3(u)-\left[\varepsilon_\alpha-\frac{1}{2}\omega^2 u^2\right]\rho(u)+\frac{\alpha}{2}\rho^2(u)\right\}\mathd u, \label{eq.Veff}
\end{equation}
where
\begin{equation}
\varepsilon_\alpha\equiv \tilde  \mu_F/N ,\quad \tilde \mu_F \equiv \mu_F+\alpha\delta(u-u).
\end{equation}
The symbol $\delta(u-u)$ could be thought of as the diagonal element of the identity operator in the coordinate representation,
\begin{equation}
\delta(u-u)=\bra u\hat I \ket u.
\end{equation}
As argued in~\cite{Douglas:1994zu}, to apply the Coleman-Luther-Mandelstam bosonization to the non-relativistic fermion, one has to extend the fermion sea to infinity. The divergent $\alpha \delta(0)/N$ term for any non-vanishing $\alpha$, though sub-leading in $N$, may provide a natural explanation for such an extension. Therefore as long as $\alpha$ is non-vanishing, the system will exhibit some kind of fermionic behavior, and the Thomas-Fermi approximation could be used~\cite{Cazalilla2011}. As $\alpha\to 0$, one would expect a phase transition to a true bosonic phase.

Doing the functional derivative with respect to $\rho(u)$, one obtains the corresponding equation
\begin{equation}
\frac{\pi^2}{2}\rho^2(u)+\alpha \rho(u)=\varepsilon_\alpha - \frac{1}{2}\omega^2 u^2,
\end{equation}
This is essentially an explicit hydrodynamic equation~\cite{Dunjko2001,Cazalilla2011}.
Despite the simple form of the above equation, it is not quite easy to solve due to the implicit dependence of $\varepsilon$ and the range of $u$ on $\alpha$. This is hidden in the normalization
\begin{equation}
\int \mathd u ~\rho(u)=1,  \label{eq.Norm2}
\end{equation}
which is inferred from (\ref{eq.Norm1}). Moreover, both the limits $\alpha\to 0$ and $\alpha\to \infty$ are not smooth and correspond to a phase transition. When $\alpha\to \infty$, the two $\alpha$ dependent terms are singled out
\begin{equation}
V_{eff}^\alpha = N^2 \int \left\{\frac{\alpha}{2}\rho^2(u) -\left[\frac{\alpha}{N}\delta(u-u)\right]\rho(u)\right\}\mathd u,
\end{equation}
Functional differential to $\rho(u)$ leads to
\begin{equation}
\rho(u)=\frac{1}{N}\delta(u-u).
\end{equation}
In other words, the corresponding density operator $\hat \rho$ satisfies
\begin{equation}
\hat \rho=\frac{1}{N} \hat 1,\quad (N \hat \rho)^2=N \hat \rho.\label{eq.rho2}
\end{equation}
In the next section we will show that this condition specify the fermionic nature of the system.
When such a constraint is satisfied, the interaction term vanishes identically. This means that an infinitely strong $\delta$-function interaction is equivalent to a fermionic Pauli exclusive force, as first shown in ~\cite{Girardeau1960}. The system in such a limit is called Tonks-Girardeau gas~\cite{Tonks1936,Girardeau1960}. (\ref{eq.rho2}) does not fix the function form of $\rho(u)$, which is determined by the distance of successive particle positions. One then has to go to the sub-leading term in $\alpha$. Repeating the procedure, one finds
\begin{equation}
\rho(u)=\frac{1}{\pi}\sqrt{2\varepsilon_\infty-\omega^2 u^2}.
\end{equation}
 Thus one recovers the semi-circle law for free fermions trapped in the harmonic potential~(\ref{eq.rho-free}). The parameter $\varepsilon_\infty$ will be fixed by the normalization (\ref{eq.Norm2}), giving $\varepsilon_\infty=\omega$. The groundstate energy is easily obtained from by performing the integration in (\ref{eq.Veff}), with the result
\begin{equation}
E_\infty=N^2\varepsilon_\infty+V_{eff}=\frac{N^2}{2} \omega.
\end{equation}
This is simply the ground state energy of $N$ spinless fermions in a harmonic potential, as it should be~\cite{Ma1027506}. One may easily generalize the above discussion to the original Lieb-Liniger model, with no confining potential. At infinite interaction, one finds $\varepsilon_\infty=\frac{\pi^2}{2}$, giving
\begin{equation}
E_\infty=\frac{\pi^2}{6}~N^2.
\end{equation}
This is indeed the ground state energy for $N$ spinless fermions at large $N$~\cite{Girardeau1960}.

The free fermion nature of the collective field theory at infinite contact interaction allows for a much simpler derivation of the Hamiltonian (\ref{eq.H-phi}). The collective field is then the one-particle density matrix element of the fermion field~\cite{Polchinski:1991uq,Minic:1991rk,Dhar:1992rs}
\begin{equation}
\phi(x)=\Psi^\dagger(x) \Psi(x).\label{eq.BF}
\end{equation}
Through bosonization for the fermion field $\Psi$, one could easily obtain the leading two terms in (\ref{eq.H-phi}) ~\cite{Douglas:1993wy,Douglas:1994zu}. As argued before, the divergent $\alpha \delta(0)/N$ term may be necessary to apply such a bosonization for non-relativistic fermion. The bosonization also suggests a 2D string theory representation of the problem~\cite{Douglas:1993wy}. Alternatively, one could deduce the dynamics of $\Psi$ directly in the large-$N$ limit.
Collective evolution of $\Psi(x)$ is simply reflected in the classical motion of the Fermi surfaces, which is determined by the corresponding Euler's equation~\cite{Polchinski:1991uq,Minic:1991rk,Dhar:1992rs}. Translating the energy of the fermion liquid back with the above formula, one immediately obtains the leading two terms in (\ref{eq.H-phi}). Since these two terms come from the kinetic part of (\ref{eq.1D-Boson}), they remains of the same form away from the free fermion point. Likewise, giving (\ref{eq.H-phi}) together with the constraint (\ref{eq.rho2}), one could retain the Euler's equation for an ideal fluid of free fermions with the following identification~\cite{Douglas:1994zu}
 \begin{equation}
 v=\partial_x \pi,\quad  P=\frac{\pi^2}{6}\phi^3(x).
 \end{equation}
 Here $v$ is the fluid velocity, and $P$ is the pressure. From this comparison one recognizes the close relation to the hydrodynamic approach, which is usually derived from the Gross-Pitaevskii equation~\cite{Cazalilla2011}. Our previous approximation by neglecting the kinetic term corresponds to the static case there, with $v=0$. And the conjugate field $\pi(x)$ should be identified with the phase operator $\hat \theta$.

When $\alpha\to 0$, it has long been known that the density function is singular~\cite{Lieb:1963rt}. Later we will show that the density goes to a $\delta$-function. If $\alpha$ is small enough, the first term in the equation can be treated as a large constant with little variation. After absorbing it into the chemical potential $\varepsilon$, one obtains the parabolic form density
\begin{equation}
\rho(u)=\frac{1}{\alpha}\left[\varepsilon_0-\frac{1}{2}\omega^2u^2\right].\label{eq.rho0}
\end{equation}
 This is exactly the result obtained with the Thomas-Fermi approximation~\cite{Cazalilla2011}. See also the hydrodynamic derivation in \cite{Dunjko2001} and the numerical confirmation in \cite{Ma1027506}. The parameter $\varepsilon_0$ can be determined from the normalization (\ref{eq.Norm2}) to be
\begin{equation}
\varepsilon_0=\left(\frac{3\alpha \omega}{4\sqrt{2}}\right)^{2/3}.
\end{equation}
The corresponding energy of the ground state is
\begin{equation}
E_0=N^2 \varepsilon_0 +V_{eff}=\frac{3}{5} N^2 \varepsilon_0=\frac{3}{5}\left(\frac{3\omega}{4\sqrt{2}}\right)^{2/3}~\alpha^{2/3}~N^2.
\end{equation}
This is exactly the leading term found in \cite{Ma1027506}, obtained in the hydrodynamic approach together with a localized Fredholm equation~\cite{Ma0926506,Dunjko2001}. The present framework therefore provides a natural basis for the validation of such a localized approach. When the harmonic potential is turned off, the chemical potential turns out to be $\varepsilon_0=\alpha$, resulting
\begin{equation}
E_0=\frac{\alpha}{2}~N^2.
\end{equation}
Again without using the Fredholm equation, which results from the Bethe ansatz~\cite{Bethe1931} at large $N$, we recover the leading term in the ground-state energy~\cite{Lieb:1963rt}~\footnote{The normalization for the energy here differs by a factor of two from ~\cite{Lieb:1963rt}.}. However, similar as the Fredholm equation, it seems not easy to go beyond the leading order analytically, especially at small $\alpha$.

Now let us take the $\alpha\to 0$ limit. It is easy to find from (\ref{eq.rho0}) $\rho(0)\sim \alpha^{-1/3}$, while the maximum value for nonzero density, $u_0$, decreases as $\alpha^{1/3}$. Taking into account of the normalization (\ref{eq.Norm2}), one concludes that as $\alpha \to 0$,
\begin{equation}
\rho(u)\to \delta(u) .\label{eq.delta}
\end{equation}
This corresponds to a transition to the Bose-Einstein condensation~(BEC) phase. In other words, the BEC phase shrinks to a single point $\alpha=0$ in the planar limit~\cite{Petrov2004,Ma1027506,Cazalilla2011}. We will get the same conclusion if we start from $N$ harmonic oscillators with no interaction, and then take the large-$N$ limit. In this case the effective potential is dominated by the limear terms~\cite{Jevicki:1979mb}
\begin{equation}
V^0_{eff}=\frac{1}{8}\int \frac{(\partial_x \phi)^2}{\phi}\mathd x+\frac{1}{2}\omega^2\int x^2 \phi(x)~\mathd x. \label{eq.V0}
\end{equation}
Due to the absence of the interaction, the chemical potential $\mu_F$ vanishes and Bose-Einstein condensation occurs. The ground state density function can be directly obtained from minimizing $V^0_{eff}$, and is given by the Gaussian function
\begin{equation}
\phi_0(x)=N\sqrt{\frac{\omega}{\pi}}\mathe ^{-\omega x^2}.
\end{equation}
Such a derivation is similar to the linearized Gross-Pitaevshii equation~\cite{Cazalilla2011}.
It is not difficult to check that when $N\to \infty$,
\begin{equation}
\phi_0(x)/\sqrt{N}\to \delta(x/\sqrt{N}),
\end{equation}
just as in eq. (\ref{eq.delta}). As a result, the $N^2$ part of the ground-state energy (\ref{eq.V0}) vanishes and the non-trivial contribution appears at ${\mathcal O} (N)$.

For intermediate $\alpha$, numerical technique is needed to find the exact density function. When the density $\rho(u)$ is obtained, we can again integrate (\ref{eq.H-phi}) to get the energy. Notice that we never deal with the momentum distribution, as studied in \cite{Lieb:1963rt,Ma0926506,Ma1027506}. It will be interesting to start from a momentum density function $\rho(k)$, and express the Hamiltonian completely in terms of $\rho(k)$. One would expect that minimizing the Hamiltonian gives rise to some generalized Fredholm equation for $\rho(k)$~\cite{Lieb:1963rt,Ma0926506,Ma1027506}. However, it seems not easy to express the $\delta$-function interaction in momentum picture directly. Perhaps the introduction of the phase space density is needed, as shown in the next section.


\section{Phase space description and 1D interacting fermions}
Now we try to extend the above formalism to the fermion system. As shown in (\ref{eq.BF}), in the infinite interaction limit the density function $\phi(x)$ be expressed through the fermion field $\Psi(x)$.
For interaction fermions it is natural to consider the following operator~\cite{Dhar:1992hr}
\begin{equation}
\hat \Phi(t)\equiv |\Psi(t)\rangle \langle \Psi(t)|,
\end{equation}
where $|\Psi(t)\rangle$ is the single-particle state vector.
In the coordinate basis, one finds
\begin{equation}
\Phi(x,y)=\Psi(x,t)\Psi^\dagger(y,t).
\end{equation}
Using the anti-commutation relation, it can be shown that $\hat \Phi(t)$ satisfies the following relation
\begin{equation}
\hat \Phi(t) ^2 =(1+N) \hat \Phi(t).\label{eq.Phi2}
\end{equation}
Employing the second quantization relation for fermions, the proper one-particle density operator could be defined as~\cite{Dhar:1992hr}
\begin{equation}
\hat \phi \equiv \hat 1-\hat \Phi,
\end{equation}
which in coordinate representation reads
\begin{equation}
\phi(x,y)=\Psi^\dagger (x,t)\Psi(y,t).
\end{equation}
From the above expression it acquires the correct normalization
\begin{equation}
\tr \hat \phi =N. \label{eq.phi-constraint2}
\end{equation}
With (\ref{eq.Phi2}) one can check that it satisfies also
\begin{equation}
\hat \phi^2 =\hat \phi, \label{eq.phi-constraint1}
\end{equation}
which represents the fermionic structure of the system. In the previous section, the operator $N \hat \rho$ in the strong-interacting limit satisfies exactly the same constraint~(\ref{eq.rho2}).
The free fermion system can be conveniently described through the coadjoint orbit/coherent state of the $W_\infty$ algebra~\cite{Dhar:1992hr}, which is the algebra of differential operators in the single-particle Hilbert space~\cite{Pope:1990rn}. Roughly speaking, under the action of the $W_\infty$ transformation, the operators follow a coadjoint orbit, and the states become coherent ones.
Such a description has a close relation with the two-dimensional string theory~\cite{Polchinski:1991uq,Minic:1991rk,Dhar:1992rs}. For non-interacting fermions in a central potential, the action can be compactly written as
\begin{equation}
S[\hat \phi]=\mathi \int \mathd s \mathd t \tr (\hat \phi [\partial_t \hat \phi,\partial _s \hat \phi])-\int \mathd t \tr (\hat \phi \hat h),
\end{equation}
where $\hat h$ is the single-particle Hamilton operator, with the element
\begin{equation}
\langle x|\hat h(t) |y\rangle\equiv h(x,y,t)=\frac{1}{2}(\partial_x ^2-V(x))\delta(x-y).\label{eq.h}
\end{equation}
Varying $\hat \phi$ along the adjoint orbit with (\ref{eq.phi-constraint1}) and (\ref{eq.phi-constraint2}) preserved, one gets the equation of motion
\begin{equation}
\mathi \partial_t \hat \phi+[\hat h,\hat \phi]=0. \label{eq.free-fermion}
\end{equation}
In the time-independent limit, the equation can be easily solved. Using the phase space representation, the equation becomes
\begin{equation}
\{ h(p,q), \phi(p,q)\}_{MB}=0,
\end{equation}
where MB denotes the Moyal bracket~\cite{Moyal1949}.
Therefore $\phi(p,q)$ should depend on the phase variables only through $h(p,q)$. Moreover, in the large-$N$ limit the constraint (\ref{eq.phi-constraint1}) simply requires
\begin{equation}
\phi(p,q)^2=\phi(p,q).\label{eq.fermion-c}
\end{equation}
 Taking all these into account the ground state density is given by
\begin{equation}
\phi(p,q)=\theta (\epsilon_F-h(p,q)),\label{eq.phi-phase}
\end{equation}
with $\epsilon_F$ the Fermi energy determined by the conservation of particle number.

It will be interesting to extend the discussion to the interacting case. In particular we want to discussion the case with a $\delta$-function interaction as in the bosonic case. Due to the Pauli exclusion principle, such a interaction makes no effects among spinless fermions. Therefore one has to introduce the spin degrees of freedom. The two-component fermion system with a $\delta$-function interaction is first studied in~\cite{Gaudin1967} and~\cite{Yang:1967bm}, and thus called the Gaudin-Yang model. Recently the exact solution when the interaction strength $g\to \infty$ is obtained in~\cite{Guan2009}. A nice review of the historic development of the model and the experimental progress is given in ~\cite{Guan2013}. The $\delta$-function interaction can be expressed through the density operator as
\begin{equation}
S_{int}=\alpha \frac{N_1 N_2}{N}\int \mathd t \tr(\hat \phi_1 \hat \phi_2),\label{eq.F-int}
\end{equation}
where $\alpha\equiv g/\sqrt{N}$, and $\hat \phi_i~(i=1,2)$ correspond to different spin directions. Here we are taking the limit that for each spin direction there are a large number of particles, $N_1$ and $N_2$ respectively, and both of them are of the same order as the total number $N$~\cite{Ma0926505}. In writing the above formula we have implicitly used the classical nature of the density operator in the large-$N$ limit,
\begin{equation}
\langle x| \hat \phi | y \rangle=\phi(x) \delta(x-y),
\end{equation}
which is simply the coordinate representation of (\ref{eq.rho2}). One can check that if $\phi_1$ and $\phi_2$ are identical, the interaction is  indeed trivial due to (\ref{eq.phi-constraint1}) and (\ref{eq.phi-constraint2}). Now with the interaction, the equations for $\hat \phi_1$ and $\hat \phi_2$ become entangled
\begin{eqnarray}
&&\mathi \partial_t \hat \phi_1+[\hat h,\hat \phi_1]=\alpha \frac{N_1 N_2}{N}[\hat \phi_2, \hat \phi_1]\nonumber\\
&&\mathi \partial_t \hat \phi_2+[\hat h,\hat \phi_2]=\alpha \frac{N_1 N_2}{N}[\hat \phi_1, \hat \phi_2].\label{eq.F1F2}
\end{eqnarray}
In the first equation the back-reaction of $\hat \phi_2$ to $\hat \phi_1$ is proportional to the finite ratio $N_2/N$. The $N_1$-factor cancels out since the corresponding Fermi energy scales as $N_1$. The conclusion is also true for the second equation. Therefore only the finite ratios remain in the large-$N$ limit, as one expects.

In the limit $\alpha\to 0$ they decouples, and one recovers two free-fermion systems. For small $\alpha$ the coupled equations could be solved perturbatively. For an infinitely repulsive interaction, minimizing the action (\ref{eq.F-int}) forces the product of the two density operators to vanish
 \begin{equation}
 \hat \phi_1 \hat \phi_2=0.
 \end{equation}
 In other words, fermions of different spin direction can not occupy the same energy state/phase space region any longer. This gives rise to an effective exclusive condition between different spin components~\cite{Girardeau1960,Guan2009}. Therefore the total density operator, $\hat \phi\equiv \hat \phi_1 +\hat \phi_2$, describes simply a free-fermion system with no distinguishing of the spin occupation. That means, $\hat \phi$ satisfies the constraints (\ref{eq.phi-constraint1},\ref{eq.phi-constraint2}) and the free equation (\ref{eq.free-fermion}), and the solution in phase space is given by (\ref{eq.phi-phase}). Notice that such a conclusion is actually independent of the exact form the single-particle Hamiltonian (\ref{eq.h}). The situation for general $g$ will be complicated. Using the phase space density $\phi(p,q)$, the equations (\ref{eq.F1F2}) can be reexpressed in terms of the Moyal bracket, which reduces to the Poisson bracket in the large-$N$ limit. Since the Poisson bracket involves only double derivative, the equations seem to be tractable. It would be interesting to see if in this way one can recover some kind of generalized Fredholm equation~\cite{Yang:1967bm,Ma1027501}.

\section{Summary}
In the paper we have shown how to obtain the planar limit of one dimensional many-body system. For the free fermion system it inherits from the matrix theory. In the bosonic case, the interactions can be introduced as the potential term of the density field, the collective field. Such a filed behaves classically in the planar limit, and can be analyzed easily. For the fermions, only interactions between different components survive. In the planar limit they can be expressed as the overlap integral of the phase density functions. Evolution of the system is determined by the corresponding classical equations in terms of Poisson brackets.

We use the $\delta$-function interaction to show how these methods can be explicitly applied. Such a model exhibits a dynamical evolution from an interacting system to a free fermionic one. This provides a dynamical bosonization of nonrelativistic fermions, and shows the differences between the collection field theory and the fermion phase description. Without using the Fredholm equation from the Bethe ansatz, we recover the correct results at strong and weak interacting limits. In this sense, the present formalism serves as a general complementary description to the Bethe ansatz, which is specific for the model. In particular, the planar limit gives a solid basis for the validation of the localized Fredholm equation proposed recently. The methods could be immediately applied to general interactions, and even higher dimensional systems. Also one could include the sub-leading terms to study the finite-$N$ corrections. One may expect such corrections give rise to the low energy fluctuations around the ground state density. We will try to investigate some of these topics in the future.

\section*{Acknowledgments}
The work is partially supported by the National Natural Science Foundation of China under Grant No. 11405065 and No. 11445001.

\end{document}